\documentclass[aps,prd,reprint,twocolumn,preprintnumbers,superscriptaddress,nofootinbib]{revtex4-2}

\usepackage{orcidlink}
\usepackage{amsmath}
\usepackage{graphicx}
\usepackage{float}
\hypersetup{colorlinks=True, citecolor=blue, urlcolor=blue, linkcolor=blue}
\usepackage[normalem]{ulem}

\begin{document}

\author{Kevin Braga\orcidlink{0009-0008-8752-8292}}
\affiliation{Department of Physics, William \& Mary, Williamsburg, Virginia 23185, USA}

\author{Markus Diefenthaler\orcidlink{0000-0002-4717-4484}}
\affiliation{Thomas Jefferson National Accelerator Facility, Newport News, VA 23606, USA}

\author{Steven Goldenberg\orcidlink{0000-0002-5264-6298}}
\affiliation{Thomas Jefferson National Accelerator Facility, Newport News, VA 23606, USA}

\author{Daniel Lersch}
\email{dlersch@jlab.org}
\affiliation{Thomas Jefferson National Accelerator Facility, Newport News, VA 23606, USA}

\author{Yaohang~\surname{Li}\orcidlink{0000-0003-0178-1876}}
\affiliation{Department of Computer Science, Old Dominion University, Norfolk, Virginia 23529, USA}

\author{Jian-Wei Qiu}
\affiliation{Thomas Jefferson National Accelerator Facility, Newport News, VA 23606, USA}

\author{Kishansingh Rajput}
\affiliation{Thomas Jefferson National Accelerator Facility, Newport News, VA 23606, USA}
\affiliation{Department of Computer Science, University of Houston, Houston, TX 77204, USA}

\author{Felix Ringer\orcidlink{0000-0002-5939-3510}}
\email{felix.ringer@stonybrook.edu}
\affiliation{Department of Physics and Astronomy, Stony Brook University, Stony Brook, NY 11794, USA}

\author{Nobuo Sato\orcidlink{0000-0002-1535-6208}}
\email{nsato@jlab.org}
\affiliation{Thomas Jefferson National Accelerator Facility, Newport News, VA 23606, USA}

\author{Malachi Schram\orcidlink{0000-0002-3475-2871}}
\affiliation{Thomas Jefferson National Accelerator Facility, Newport News, VA 23606, USA}

\preprint{JLAB-THY-25-4413}

\date{\today}

\title{Toward an event-level analysis of hadron structure using differential programming}

\begin{abstract}
Reconstructing the internal properties of hadrons in terms of fundamental quark and gluon degrees of freedom is a central goal in nuclear and particle physics. This effort lies at the core of major experimental programs, such as the Jefferson Lab 12 GeV program and the upcoming Electron-Ion Collider. A primary challenge is the inherent inverse problem: converting large-scale observational data from collision events into the fundamental quantum correlation functions (QCFs)  that characterize the microscopic structure of hadronic systems within the theory of QCD. Recent advances in scientific computing and machine learning have opened new avenues for addressing this challenge using deep learning techniques. A particularly promising direction is the integration of theoretical calculations and experimental simulations into a unified framework capable of reconstructing QCFs directly from event-level information. In this work, we introduce a differential sampling method called the local orthogonal inverse transform sampling (LOITS) algorithm. We validate its performance through a closure test, demonstrating the accurate reconstruction of a test distribution from sampled events using Generative Adversarial Networks. The LOITS algorithm provides a central building block for addressing inverse problems involving QCFs and enables end-to-end inference pipelines within the framework of differential programming.
\end{abstract}
\maketitle
\tableofcontents

\section{Introduction}
\label{s.intro}

The next generation of experiments in nuclear particle physics, including the ongoing Jefferson Lab 12 GeV program and the future Electron-Ion Collider (EIC), is expected to deliver an unprecedented volume of data for studying the internal properties of hadrons and nuclei. One of the primary scientific endeavors is to reconstruct the internal three-dimensional structure of nucleons and nuclei in terms of their quark and gluon degrees of freedom. These structures are encoded in quantum correlation functions (QCFs), such as parton distribution functions (PDFs), transverse momentum distributions (TMDs), and generalized parton distribution functions (GPDs), all of which are defined within the fundamental theory of strong interactions. While PDFs describe the longitudinal momentum distribution of partons inside hadrons, TMDs and GPDs account for the transverse momentum and spatial distribution, respectively. Reconstructing such quantities presents an inverse problem, as they cannot be directly measured in experiments. Instead, they need to be inferred from observational data that consist of large collections of collision events, each composed of data structures for the final-state particles.

Traditionally, QCFs are inferred from differential cross sections that are derived from event-averaged, binned quantities, also referred to as summary statistics. These cross sections are extracted from
measured data using an unfolding procedure based on simulations. The unfolding accounts for model uncertainties, detector effects, and background
contributions. A commonly used technique for this purpose is Iterative Bayesian Unfolding~\cite{DAGOSTINI1995487}. As a second step, QCFs are extracted from the unfolded data using QCD factorization theorems~\cite{Collins:1989gx}, which separate the cross section into perturbatively calculable components and the nonperturbative QCFs of interest. Constraints from lattice QCD can also be incorporated~\cite{Lin:2017stx, Bringewatt:2020ixn, JeffersonLabAngularMomentumJAM:2022aix, Cocuzza:2023oam, Gamberg:2022kdb, Karpie:2023nyg, Hunt-Smith:2024khs}. See Refs.~\cite{Ablat:2024muy, 
NNPDF:2024dpb, Bhattacharya:2021twu, Cridge:2024icl, 
Barry:2021osv, Cocuzza:2021rfn, Hunt-Smith:2024khs, Cocuzza:2025qvf, Borsa:2024mss, Borsa:2021ran, Guo:2025jiz, Bacchetta:2025ara, Moos:2025sal, Barry:2023qqh, Cruz-Martinez:2025ahf, AbdulKhalek:2022laj} for examples of this approach.

Looking ahead, to reduce the uncertainties on QCFs and sharpen our understanding of nucleon and nuclear structure, full end-to-end simulations may become feasible given recent advances in AI and machine learning~\cite{Boehnlein:2021eym}. These could make use of event-level data without loss of information. This approach is expected to be particularly valuable for extracting the three-dimensional structure of nucleons and nuclei, which needs to be inferred from multi-differential cross sections. In general, such an event-level analysis involves the following steps:
i) Proposing a set of trial QCFs based on a well-motivated modeling strategy;
ii) Converting the QCFs into differential cross sections via QCD factorization theorems;
iii) Generating phase-space or event-level samples from the resulting cross sections;
iv) Simulating realistic experimental data, including detector effects and backgrounds, using the generated events from iii); and
v) Optimizing the QCFs by minimizing a distance metric between the simulated and observed event samples.
To implement this approach in practice, a fully differential simulation and analysis pipeline is required. In the following, we discuss various components of this strategy in more detail. 

{\bf Event-level data.} In recent years, several efforts have been made to address inference problems in nuclear and particle physics using unbinned data. See, in particular, the proposal to publish unbinned data by experimental collaborations and the discussion of potential applications in Ref.~\cite{Arratia:2021otl}. In general, using unbinned data helps preserve the full information content of the dataset, avoiding information loss due to binning. For example, Ref.~\cite{Benato:2025rgo} demonstrated advantages of unbinned maximum likelihood methods applied to simulated Higgs data, showing improved performance compared to traditional binned approaches. While for low-dimensional observables an optimal, inference-aware binning can be constructed \cite{Kaastra:2016qwt, DeCastro:2018psv, Dannheim:2009zz, TMVA:2007ngy}, this becomes significantly more challenging for multi-differential observables and/or low-statistics regimes, which are particularly relevant in studies of 3D hadron structure.

{\bf Event-level theory.} To carry out unbinned inference tasks at the event level, both experimental data and theoretical predictions need to be available at the event level. For inference tasks involving parton showers~\cite{Brehmer:2018kdj,Cranmer:2019eaq,JETSCAPE:2020mzn,Mastandrea:2024irf,Assi:2025ibi,Cheng:2025ewj} or specific fixed-order calculations~\cite{Frixione:2007vw,NNLOJET:2025rno,Borsa:2022cap}, event-level theoretical results are readily available. However, much of the current QCD phenomenology targeting the extraction of QCFs relies on analytical perturbative QCD calculations at fixed order matched to e.g. TMD resummation. In addition, theoretical calculations of exclusive processes sensitive to GPDs exist only as analytical expressions. For processes such as Deeply Virtual Compton Scattering, these results can be converted into event-level samples using algorithms like the one developed in Ref.~\cite{Berthou:2015oaw,Aschenauer:2025cdq}. See also Refs.~\cite{Perez:2004ig,Toll:2013gda,Lomnitz:2018juf,Gryniuk:2020mlh}. Since theoretical simulations are generally not differentiable, surrogate models need to be introduced to incorporate them into end-to-end inference pipelines~\cite{Bellagente:2020piv,Nachman:2022jbj,Heimel:2022wyj,Ngairangbam:2023cps,Chan:2023ume,Bahl:2024gyt,Heimel:2024wph,Araz:2025ezp}. In this work, we propose a complementary method to convert analytical cross section results into differentiable event samples, thereby avoiding the need for surrogate models. Specifically, we introduce a Local Orthogonal Inverse Transform Sampling (LOITS) algorithm that enables us to build a differentiable manifold connecting phase space samples of events with the underlying QCF model parameters.

{\bf Detector effects.} As noted above, comparing experimental data with theoretical predictions requires a careful treatment of detector effects and backgrounds. This can be achieved either by unfolding the unbinned experimental measurements from detector level to particle level or by forward-folding the theoretical predictions through a detector simulation such as \textsc{GEANT4}~\cite{GEANT4:2002zbu}. In recent years, event-level unfolding techniques have been developed to go beyond traditional methods that rely on binned data. For instance, density-based approaches have been introduced in Refs.~\cite{ Datta:2018mwd,Bellagente:2019uyp,Howard:2021pos,Bellagente:2020piv,Vandegar:2020yvw,Howard:2021pos,Backes:2022sph,Du:2024gbp,Diefenbacher:2023wec,Desai:2024yft,Butter:2024vbx,Favaro:2025psi}, while classifier-based methods for reweighting events have been proposed in Refs.~\cite{Andreassen:2019cjw,Andreassen:2021zzk,Pan:2024rfh,Milton:2025mug,Falcao:2025jom}. Some of these techniques have already been adopted by experimental collaborations, as seen in Refs.~\cite{H1:2021wkz,ATLAS:2024xxl,ATLAS:2024jry}. For certain low-energy scattering cross sections, such as those measured at Jefferson Lab, unfolding can be more challenging due to the lack of sufficiently accurate theoretical models for generating phase space samples, in contrast to the high-energy collider environment. When forward folding of theoretical predictions is used within the inference pipeline, a differentiable detector simulation is required that can be achieved using surrogate models. See, Ref.~\cite{Krause:2024avx} and references therein. Instead, when using unfolded event samples, correlations between events need to be accounted for~\cite{Desai:2025mpy}. 

{\bf Parametrization of QCFs.} The modeling of QCFs introduces an additional computational challenge. Strictly speaking, QCFs are continuous functions, and in general, it is not possible to constrain them directly without preconditioning the problem, typically by truncating their local analytic behavior. In practice, a variety of parameterizations are routinely employed. These range from simple functional forms involving ${\cal O}(10-100)$ parameters~\cite{Gamberg:2022kdb} to more flexible representations using neural networks~\cite{NNPDF:2017mvq,Bacchetta:2025ara}, which typically involve a significantly larger number of parameters. As the complexity of QCFs increases, for example, for multidimensional structures such as GPDs, more flexible parameterizations are needed to mitigate modeling bias and reliably estimate uncertainties. Multidimensional parton densities are well-suited to be modeled as pixelated images, allowing for a direct quantification of the resolution at which QCF images can be reconstructed from data. As an illustrative example, in this work, we consider Generative Adversarial Networks (GANs), which can be trained on event-level data to generate pixelated QCF images that are consistent with observational constraints. While this example application is based on the application of generative AI, we note that the LOITs sampling algorithm is independent of this particular approach.

{\bf Uncertainties.} Unlike fundamental physical parameters such as the $W$-boson mass, electroweak couplings, or Standard Model Effective Field Theory parameters, model parameters used in QCF modeling are subject to priors that regularize the inverse problem~\cite{DelDebbio:2021whr}. Uncertainty quantification for QCFs has a long history, beginning with the pioneering work of the CTEQ collaboration~\cite{Pumplin:2002vw}. Subsequently, the NNPDF collaboration introduced neural network–based parametrizations and employed ensemble-based approaches to address the uncertainty quantification for PDFs. Since then, various groups have adopted ensemble methods for a variety of QCFs, including helicity-dependent PDFs, fragmentation functions, TMDs, and GPDs~\cite{Sato:2016tuz, Sato:2016wqj, Bacchetta:2024qre, Guo:2025jiz}. Generally, these approaches fundamentally rely on observational data presented as summary statistics (e.g., differential cross sections, asymmetries), and to our knowledge, a fully realized framework for reconstructing QCFs with uncertainty quantification carried out directly at the event level has not yet been achieved. Nevertheless, ensemble-based methodologies can, in principle, be extended to event-level QCF inference through bootstrap resampling of events combined with the propagation of uncertainties arising from unfolding or folding procedures. While a comprehensive treatment of the uncertainty quantification for QCF reconstruction at the event level is beyond the scope of this work, our focus in this document is to provide a differentiable solution for realizing end-to-end, simulation-based inference that can leverage generative AI techniques. Since we are particularly interested in multidimensional QCFs represented as images, we will also discuss a specific aspect of UQ related to the resolution of the images that our trained models can infer.

The remainder of this document is organized as follows: In Section~\ref{s.physics}, we briefly review the underlying physics of QCFs, their connection to observational data, and relevant aspects of detector simulations. In Section~\ref{s.loits}, we introduce the differentiable sampling algorithm LOITS. Section~\ref{s.Test case} presents a test case demonstrating the performance of our approach using GANs. Finally, we conclude in Section~\ref{s.conclusions}.

\section{Physics of Quantum Correlation Functions} 
\label{s.physics}

QCFs are nonperturbative objects defined within QCD that characterize the internal structure of hadrons and nuclei in terms of their quark and gluon degrees of freedom. They are typically defined as matrix elements of QCD field operators evaluated between hadronic or nuclear states. A classical example of QCFs are collinear PDFs, which have a long history in high-energy and nuclear particle physics. These objects represent the number densities of partons carrying a fraction of the large light-cone momentum of their parent hadron. Using QCD factorization theorems, one can compute differential cross sections for a given process as a convolution of QCFs with process-dependent parton-level cross sections that are calculable in perturbative QCD. This framework enables the description of a wide range of processes at different collider experiments.

While PDFs are one-dimensional QCFs, there exist extensions that encode the three-dimensional structure of hadrons. For instance, transverse momentum-dependent (TMD) PDFs are three-dimensional QCFs that include both the longitudinal momentum fraction and the intrinsic nonperturbative transverse momentum distribution of partons inside hadrons, offering a momentum-space tomography. Complementary to this are Generalized Parton Distributions (GPDs), which are three-dimensional QCFs encoding the spatial distributions of partons, thereby enabling position-space tomography of hadrons or nuclei. 

To access the spatial distribution of partons inside a proton, we can study for example the following exclusive scattering process $ep \to e'p'\gamma$, which allows for the reconstruction of so-called Compton Form Factors (CFFs) in deeply virtual Compton scattering (DVCS) \cite{Ji:1996nm}. These CFFs play a role analogous to that of structure functions in Deep Inelastic Scattering (DIS) and can be factorized in terms of GPDs and short-distance partonic coefficient functions:
\begin{align}
{\cal F}^a_N(\xi,t,Q^2) = \int_{-1}^{1} dx \sum_{i} C^a_i(x,\xi,Q^2;\mu^2)
F^a_{i/N}(x,\xi,t;\mu^2)\,.
\label{eq.gpds}
\end{align}
Here, ${\cal F}^a_N$ denotes a CFF, expressed in terms of the momentum transfer to the target nucleon $t$, the longitudinal momentum transfer $\xi$, and the virtuality of the exchanged photon $Q^2$. The index $a$ distinguishes different types of CFFs associated with the polarization of the initial-state nucleon, as well as the possible spin-conserving and spin-flip transitions of the diffracted final state~\cite{Ji:1996nm, Belitsky:2005qn}. For spin-$\tfrac{1}{2}$ particles such as nucleons, there are four independent CFFs, each corresponding to one of the four associated GPDs, denoted $F^a_{i/N}$, where the additional index $i$ labels the parton flavor. Similar to PDFs, GPDs depend additionally on the parton's light-cone momentum fraction $x$ and the factorization scale $\mu^2$ governed by GPD evolution equations~\cite{Ji:1996nm, Freese:2024ypk, Vinnikov:2006xw, Bertone:2022frx}. The CFFs are then obtained by integrating the GPDs against perturbatively calculable coefficient functions $C^a_i$. 

In contrast to structure functions in DIS, the formulation of CFFs in DVCS requires integration over the full range of the parton momentum fraction, $-1 < x < 1$. This presents significant challenges for reconstructing the $x$-dependence of GPDs \cite{Bertone:2021yyz, Moffat:2023svr}, as the reconstruction is effectively limited to the kinematic ridge at $x = \xi$. However, complementary observables that provide constraints away from the ridge do exist, such as exclusive photoproduction of back-to-back $\pi,\gamma$ system \cite{Qiu:2023mrm}.

The multidimensional nature of GPDs demands careful modeling strategies in inference tasks. For example, the Fourier transform of GPDs with respect to $t$ provides access to the coordinate-space density of partons relative to the hadron’s center, offering a spatial image of parton distributions. In this context, quantifying the resolution at which experimental data can constrain these coordinate-space GPD densities becomes particularly important. Similar considerations apply not only to GPDs, but also to TMDs and even collinear PDFs. From this perspective, machine learning models for QCFs offer a promising strategy for addressing inverse problems, particularly given their flexibility in optimizing large parameter spaces. While the numerical techniques developed in the following sections are broadly applicable to inverse problems in nuclear and particle physics, our primary area of focus is nuclear imaging involving multidimensional QCFs, such as GPDs or TMDs.

\section{Differentiable sampling algorithm} 
\label{s.loits}

\begin{figure}[t]
    \centering   
    \includegraphics[trim={0 1.0cm 10cm 0.1cm},clip,width=0.4\textwidth]{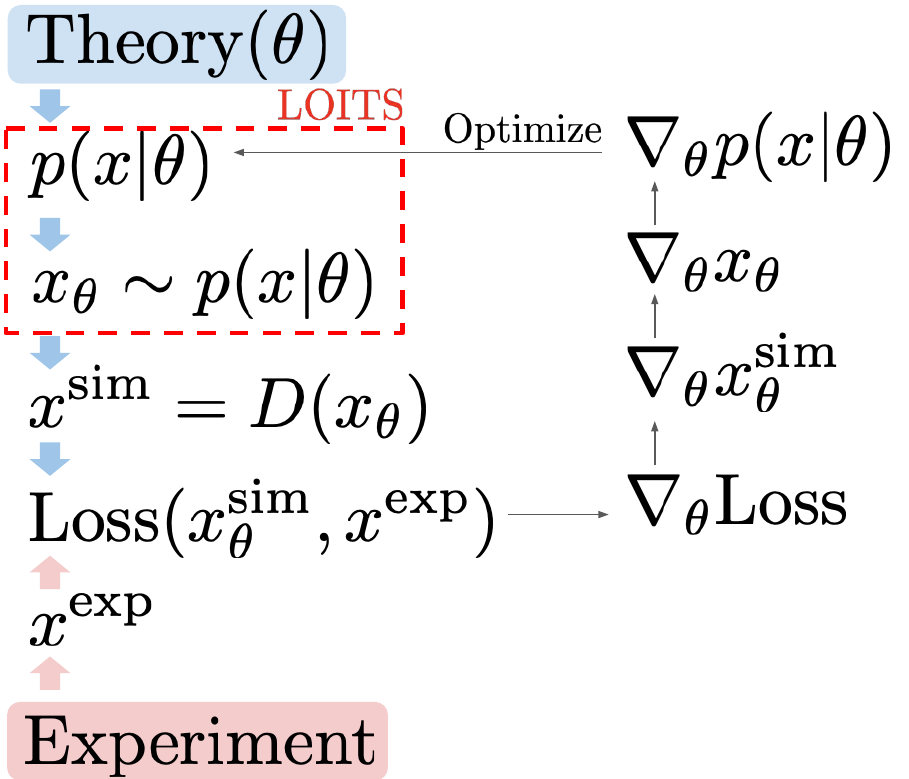}
    \caption{Schematic illustration of the simulation-based inference pipeline and the backpropagation chain for optimizing QCF model parameters. Phase space samples \( x_{\theta} \) are drawn from a theoretical density \( p(x|\theta) \), conditioned on the QCF model parameters \( \theta \). These samples are passed through a detector simulation $D$ to produce \( x_{\theta}^{\rm sim} \), which are compared to experimental data \( x^{\rm exp} \) using a loss function that quantifies their discrepancy. The LOITS algorithm introduced in this work enables a fully differentiable sampling procedure, creating a continuous and differentiable path between \( \theta \) and the loss. The red box highlights the differentiable sampling stage, enabling backpropagation through the entire pipeline.}
    \label{f.workflow}
\end{figure}

As discussed above, this work aims to provide a solution for QCF model optimization using event-level data. Our goal is to construct a differentiable map between phase space samples and QCF model parameters. The proposed strategy is rather general and can be applied to simple parametric forms of QCFs, neural network parametrizations, as well as generative models where QCFs are represented as images.

Let us denote by $p(x|\theta)$ a theoretical description of the phase space density, conditioned on a set of tunable QCF model parameters $\theta$, where $x$ denotes the phase space features.  This quantity can be constructed using differential cross sections derived from QCD factorization theorems, as discussed in section~\ref{s.physics}. For example, in exclusive processes involving the DVCS subprocess, one has $p(\xi,t,Q^2|\theta)\propto d\sigma/d\xi\,dt\,dQ^2$ where $\theta$ denotes the input model parameters of the GPDs. With this in mind, the simulation pipeline can be schematically written as
\begin{align}
\theta
\to p(x|\theta)
\to x_{\theta}
\to D(x_{\theta})
\to x^{\rm sim}_{\theta} \,.
\end{align}
Here, the samples $x_{\theta}$ are drawn from the theoretical phase space density, $x_\theta\sim p(x|\theta)$. The subscript of $x_{\theta}$ indicates that gradients $\nabla_\theta x_\theta$ need to be tractable, following the differential programming paradigm, to optimize model parameters. The generated samples can then be passed through a detector simulation $D$, which transforms $x_{\theta}$ into realistic samples $x_\theta^{\rm sim}$ that can be compared with experimental data at the event level. Typically, detector effects require simulators such as ~\cite{Agostinelli:2002hh}. Since detector simulation is generally non-differentiable, a machine learning surrogate is often needed to enable differentiability. However, in this work, we do not address this aspect and focus only on the differentiability of the phase space samples prior to detector effects. A schematic representation of the simulation pipeline is shown in Fig.~\ref{f.workflow}. The diagram illustrates the computational path connecting the QCF parameters $\theta$ to a generic loss function that quantifies the discrepancy between simulated event samples and experimental data, along with the chain of differentiability required for parameter optimization. The goal of the LOITS algorithm introduced here is to enable the computation of exact gradients of the loss function with respect to $\theta$ by constructing a differentiable map between $x_{\theta}$ and $\theta$, as highlighted by the red dashed box in Fig.~\ref{f.workflow}.  Without the differential sampling algorithm, the chain rule originating from the loss function is broken, and the gradients needed to optimize the model parameters cannot be computed. While gradient estimation via finite differences is a possible alternative, our proposed solution leverages the automatic differentiation capabilities of modern machine learning libraries to compute gradients both accurately and efficiently.

We now proceed to discuss our proposed LOITS algorithm, which enables us to build a differentiable map between the phase space samples $x_{\theta}$ and the parameters $\theta$. The focus of LOITS is the sampling procedure, indicated by the red dashed box in Fig.~\ref{f.workflow}. The key idea is that the algorithmic procedure for drawing samples from the density must retain information about the QCF parameters $\theta$. For example, when using standard Markov Chain Monte Carlo (MCMC) methods to generate samples from a target density with Gaussian proposals, the gradients vanish $\nabla_{\theta} x_{\theta} = 0$. This is because the samples are obtained from a Gaussian proposal function that has no analytical dependence on the target density's model parameters $\theta$. Instead, a differentiable sampling algorithm can be obtained using the well-known method of {\it inverse transform sampling} (ITS). To illustrate the procedure, let us first consider a simple example where the phase space $x$ is one-dimensional and restricted to the interval $0 < x < 1$\footnote{The range of support can be extended by modifying the integration limits of the CDF.}. In this case, we can calculate the cumulative density function (CDF) as
\begin{align}
    {\rm CDF}(x,\theta) = \int_0^x dz~p(z|\theta) \;.
\end{align}
By construction, ${\rm CDF}(x,\theta)$ is analytic in $\theta$ provided that $p(x|\theta)$ is itself an analytic function of $\theta$, which is typically the case for applications to QCF. Therefore, the gradient $\nabla_{\theta} {\rm CDF}$ is tractable. Applying the chain rule, we obtain
\begin{align}
    \frac{\partial {\rm CDF}(x,\theta)}{\partial \theta_i} = 
    \int_0^x dz
    \frac{\partial {\rm CDF}(x,\theta)}{\partial p(z|\theta)}
    \frac{\partial p(z|\theta)}{\partial \theta_i}\,,
\end{align}
where the subscript $i$ refers to the $i^{\rm th}$ model parameter. To perform ITS, we need to access the inverse CDF. Samples from $p(x|\theta)$ can then be generated as
\begin{align}
    \label{def_inv_cdf}
    x_{\theta}(u) = {\rm CDF}^{-1}(u,\theta)\,,
\end{align}
where $u\sim \mathcal{U}[0,1]$ is drawn from a uniform distribution over $0 < u < 1$. The resulting samples $x_{\theta}$ are then distributed according to the target density, i.e., $x_{\theta} \sim p(x|\theta)$. Importantly, for any value of $u$, the sample $x_{\theta}$ is an analytic function of $\theta$. Thus, using the chain rule, we can compute its gradient as
\begin{align}
    \frac{\partial x_{\theta}}{\partial \theta_i} 
    &= 
    \frac{\partial {\rm CDF}^{-1}(u,\theta)}{\partial \theta_i} 
    \notag\\
    &= 
    \frac{\partial {\rm CDF}^{-1}(u,\theta)}{\partial {\rm CDF}(x_{\theta},\theta)} 
    \frac{\partial {\rm CDF}(x_{\theta},\theta)}{\partial \theta_i} \; .
    \label{e.dxdtheta}
\end{align}
It is important to note that, by construction, the endpoints of the CDF are fixed for all values of $\theta$, i.e., ${\rm CDF}(0,\theta) = {\rm CDF}(1,\theta) = 1$. As a result, the gradients in Eq.~\eqref{e.dxdtheta} vanish at the endpoints $x_\theta = 0$ and $x_\theta = 1$. This implies that samples near the boundaries lose sensitivity to the model parameters $\theta$. We will discuss this aspect in more in section~ \ref{s.Test case}.

In general, it is not possible to obtain an analytical expression for ${\rm CDF}^{-1}(u,\theta)$ for an arbitrary density $p(x|\theta)$. However, provided that the density does not vanish within the phase space boundaries, which is the case for the applications considered in this work, the CDF is a strictly monotonic function. As a result, its inverse is uniquely defined. Using this feature, we can evaluate the CDF on a discrete set of values $0 < x < 1$ and construct an approximate inverse $\rm CDF^{-1}$ using a local interpolation (LI) function\footnote{We define a local interpolation as
$$y={\rm LI}(x;X,Y)$$
where $X,Y$ are pairwise connected arrays to be interpolated, and $y$ is the interpolated value at the coordinate $x$.}. 
Specifically, let $X$ be the grid of $x$ values and $Y(\theta)$ the corresponding array of CDF values, i.e., $Y(\theta)={\rm \{CDF}(x,\theta) ~{\rm for}~x~{\rm in}~X\}$. Then, samples can be approximately generated as 
\begin{align}
    x_{\theta}=
    {\rm CDF}^{-1}(u,\theta)
    &\approx
    {\rm LI}(u;X,Y(\theta)) \,.   
    \label{e.LI}
\end{align}
Provided that the interpolation function LI is analytic in $Y$, and hence in $\theta$, the samples $x_{\theta}$ generated via Eq.~\eqref{e.LI} are differentiable with respect to $\theta$. In our numerical studies presented below, we use a simple local linear interpolation function, which satisfies the required analytic condition.

\begin{figure}[t]
    \centering   
    \includegraphics[trim={0 0cm 10cm 0cm},clip,width=0.4\textwidth]{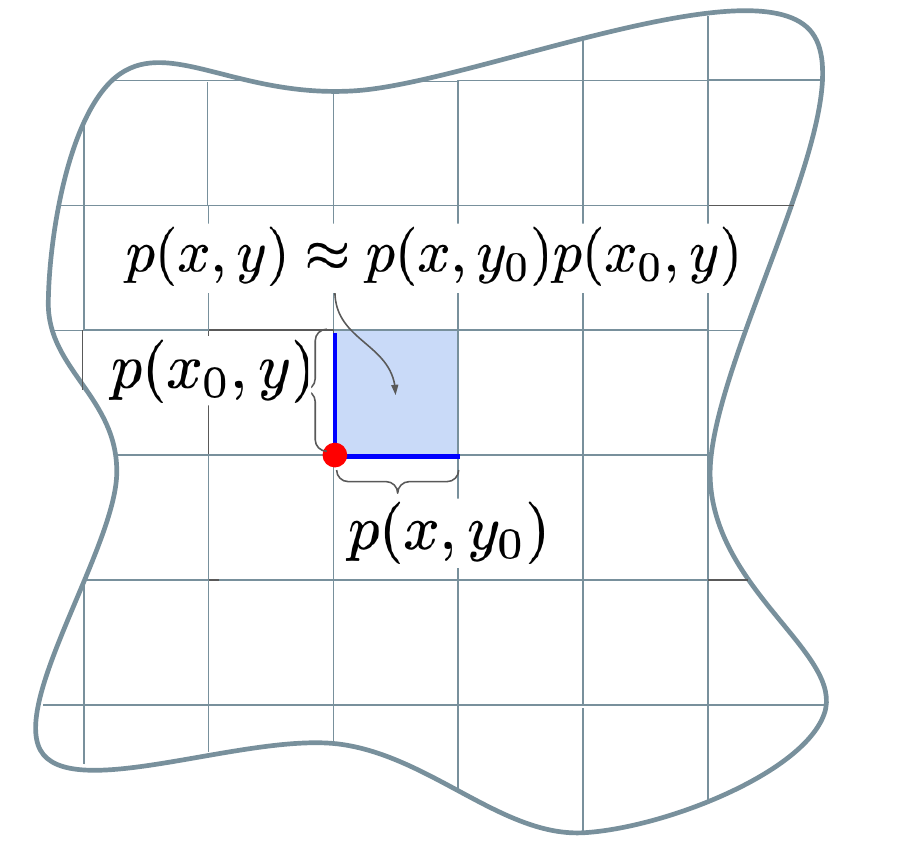}
    \caption{Illustration of the 2D phase space segmentation for implementing the LOITS algorithm. The blue lines indicate the orthogonal sampling directions used by LOITS to define local one-dimensional densities. For simplicity, the explicit dependence of the density on the model parameters \( \theta \) is omitted.}
    \label{f.2dloits}
\end{figure}
\begin{figure*}[t]
    \centering
    \includegraphics[width=1\textwidth]{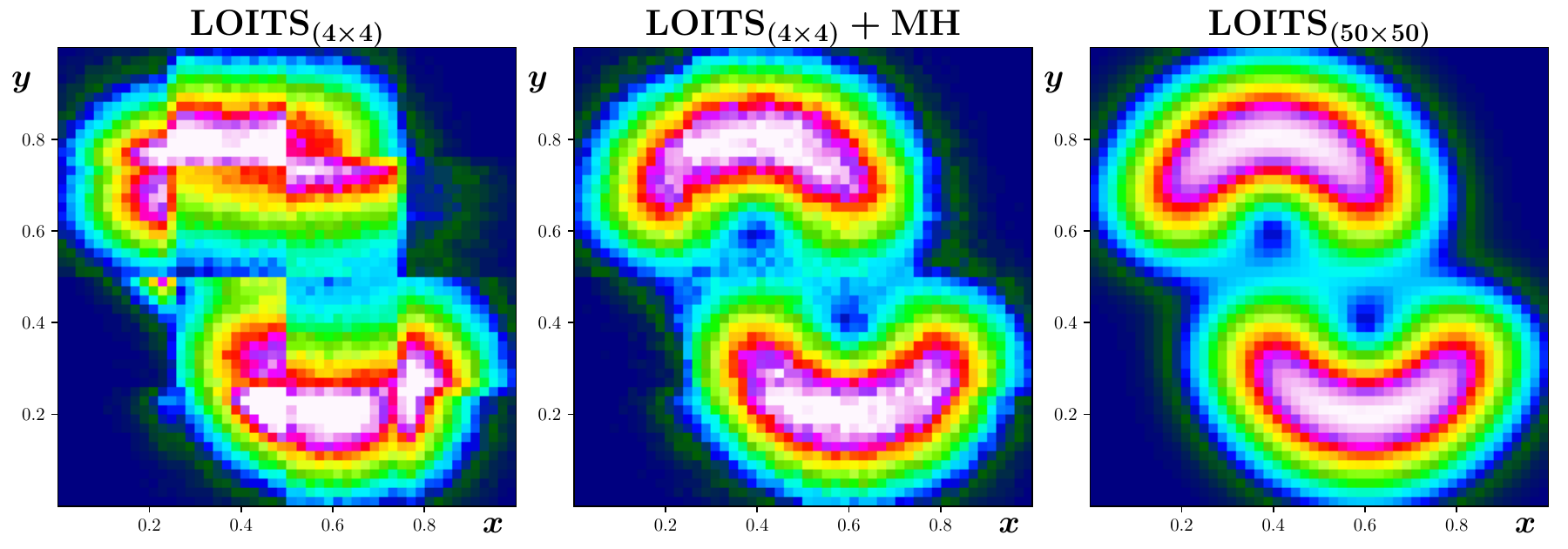}
    \caption{Sampling of the half moon distribution. \textbf{Left:} Reconstructed distribution using the LOITS algorithm with a $4\times4$ phase space segmentation.
    \textbf{Middle:} Same as the left panel, but using the MH improved LOITS algorithm. 
    \textbf{Right:} Same as the left panel, but with a $50\times50$ phase space segmentation. 
    }
    \label{fig:half_moons}
\end{figure*}

At this stage, we have demonstrated how ITS can be used to generate differentiable samples for a simplified one-dimensional setting. Unfortunately, ITS cannot be directly extended beyond one dimension because CDFs are not uniquely defined in higher dimensions. However, given that the density $p(x|\theta)$ does not vanish within the physical phase space boundaries, as is the case for the applications considered in this work, we will show that it is still possible to generalize this method to higher dimensions by introducing a local ITS algorithm. To achieve this, we begin by extending the 1D ITS method to a local sampling procedure, which can then be generalized to a higher-dimensional phase space as well. In the 1D case, we partition the phase space into $K$ segments and apply ITS within each segment independently. Let us illustrate the procedure for the 1D case introduced earlier. The steps are as follows:
\begin{enumerate}
    \item Divide the domain $0 < x < 1$ into $K$ contiguous segments: 
        \begin{align}
            \{x_k,x_{k+1}; k=1,\ldots,K\}\;.
        \end{align}
        For now, let us assume an equidistant segmentation.
    \item Compute segment probabilities and local densities: 
        \begin{align}
            m_k &=\int_{x_k}^{x_{k+1}}~dx~p(x|\theta)\,,
            \\
            p_k(x|\theta)&= \frac{1}{m_k}~p(x|\theta)~\Theta(x_k<x<x_{k+1})\;.
        \end{align}
    
    \item Compute the local CDFs for a subgrid  $x_k<x_l<x_{k+1}$ with $l=1,\ldots,L$ for all the $k$ segments, i.e.
        \begin{align}
            {\rm CDF}_k(x_i,\theta)=\int_{x_k}^{x_l} dx~p_k(x|\theta)\;.
        \end{align}

    \item To generate a total of $N$ samples, estimate the number of samples that need to be generated for each segment via
        \begin{align}
            n_k={\rm int}(m_k\cdot N)\,,
        \end{align}
    where $\mathrm{int}(\,\cdot\,)$ denotes the integer rounding operation.
    \item For each segment, draw $n_k$ uniform samples $u\in [0,1]$ and map them to phase space values $x_{\theta}$ using a local interpolation of the CDF, i.e.
        \begin{align}
            x_{\theta}
            &={\rm LI}(u; X_k,Y_k(\theta)) \;.
        \end{align}
    Here, $X_k,Y_k(\theta)$ represent the subgrid and corresponding local CDF values.
    \item Aggregate the generated samples from all the $K$ segments. 
\end{enumerate}

By construction, the collection of samples from all segments is differentiable with respect to $\theta$. Similar to the case without local sampling, the samples at the boundaries of each segment have vanishing gradients, since their corresponding local CDF values are independent of $\theta$. We revisit this aspect in detail in section~\ref{s.Test case}. 

It is important to note that the local ITS algorithm offers no advantage in the 1D case, as differentiable sampling can be accomplished without segmenting the phase space. However, the method becomes crucial when extending the algorithm to multidimensional densities. To illustrate this, consider a two-dimensional density $p(x, y|\theta)$ supported on the domain $0 < x, y < 1$, from which we wish to draw samples $(x, y) \sim p(x, y|\theta)$ such that the gradients $\nabla x_{\theta}$ and $\nabla y_{\theta}$ are tractable. We extend the local 1D ITS approach by segmenting the two-dimensional phase space and approximating the target density as a product of one-dimensional densities: $p(x, y|\theta) \approx p(x, y_0|\theta)p(x_0, y|\theta)$. Here, $x_0,y_0$ are fixed values as illustrated in Fig.~\ref{f.2dloits}. We select one coordinate to define orthogonal sampling segments, each of which defines a local one-dimensional density. Samples are then generated via ITS along each direction and concatenated to form the desired two-dimensional samples. This procedure naturally generalizes to arbitrary dimensions. Due to its local and orthogonal structure in phase space, we refer to this method as the LOITS algorithm.

While LOITS is a differentiable sampling algorithm, it is important to note that the phase space sampling is only approximate. In particular, if the target density varies rapidly, the distribution of samples generated by LOITS may not accurately reflect the true density. This issue can be mitigated by increasing the number of segments used in the local interpolation. However, this becomes less practical as the dimensionality of the phase space increases. To address this potential shortcoming, we can incorporate the Metropolis–Hastings (MH) algorithm into the sampling procedure, which leads to asymptotically exact results. We follow a strategy similar to those used for improving sampling from multidimensional distributions using normalizing flows and diffusion models~\cite{Albergo:2019eim, pmlr-v151-stimper22a, Hunt-Smith:2023ccp}. Specifically, we construct a Markov Chain Monte Carlo (MCMC) sampler using an independent proposal function based on the LOITS algorithm. Unlike traditional MCMC with Gaussian proposals, in this case the proposed states are drawn from the LOITS-generated distribution, which retains differentiability. The LOITS samples are then subjected to an additional accept–reject step using the MH criterion, which improves the sampling fidelity.

As an example, we again consider a two-dimensional phase space. A LOITS-generated state $(x, y)$ is associated with an approximate density $p^*(x, y|\theta) = p(x, y_0|\theta)p(x_0, y|\theta)$, where $(x_0, y_0)$ is the coordinate at the origin of the local orthogonal segment associated with the generated state, see Fig.~\ref{f.2dloits}. The MH acceptance probability for a newly proposed state $(x_{i+1}, y_{i+1})$ is then given by
\begin{align}
    A=\min\left[u,\frac{p(x_{i+1},y_{i+1}|\theta)}{p(x_{i},y_{i}|\theta)}
                  \cdot\frac{p^*(x_{i},y_{i}|\theta)}{p^*(x_{i+1},y_{i+1}|\theta)}
        \right]\,,
\end{align}
where $u \sim \mathcal{U}[0,1]$, and $p(x, y)$ denotes the true target density. In Fig.~\ref{fig:half_moons}, we illustrate the application of LOITS to a two-dimensional ``half-moon'' distribution. See Appendix~\ref{app:halfmoon} for a definition of the test distribution used here. In the left panel, we show the reconstructed histogram from LOITS samples with a $4 \times 4$ phase space segmentation. As expected, the result exhibits noticeable distortions, although the global features of the distribution roughly capture the two half-moon structures. Using the same setup, the middle panel shows the reconstructed histogram when LOITS is augmented with the MH step. The MH step corrects the sampling procedure, and the distribution of the two half-moons is more accurately reconstructed, thanks to the accept–reject correction of the LOITS proposals. Lastly, in the right panel, we show the results using LOITS with a finer $50 \times 50$ phase space segmentation. As expected, the accuracy improves significantly, but at the expense of more segments and a higher computational cost.

\begin{figure*}
    \centering
    \includegraphics[trim={0 3.5cm 0cm 2cm},clip,width=0.9\textwidth]{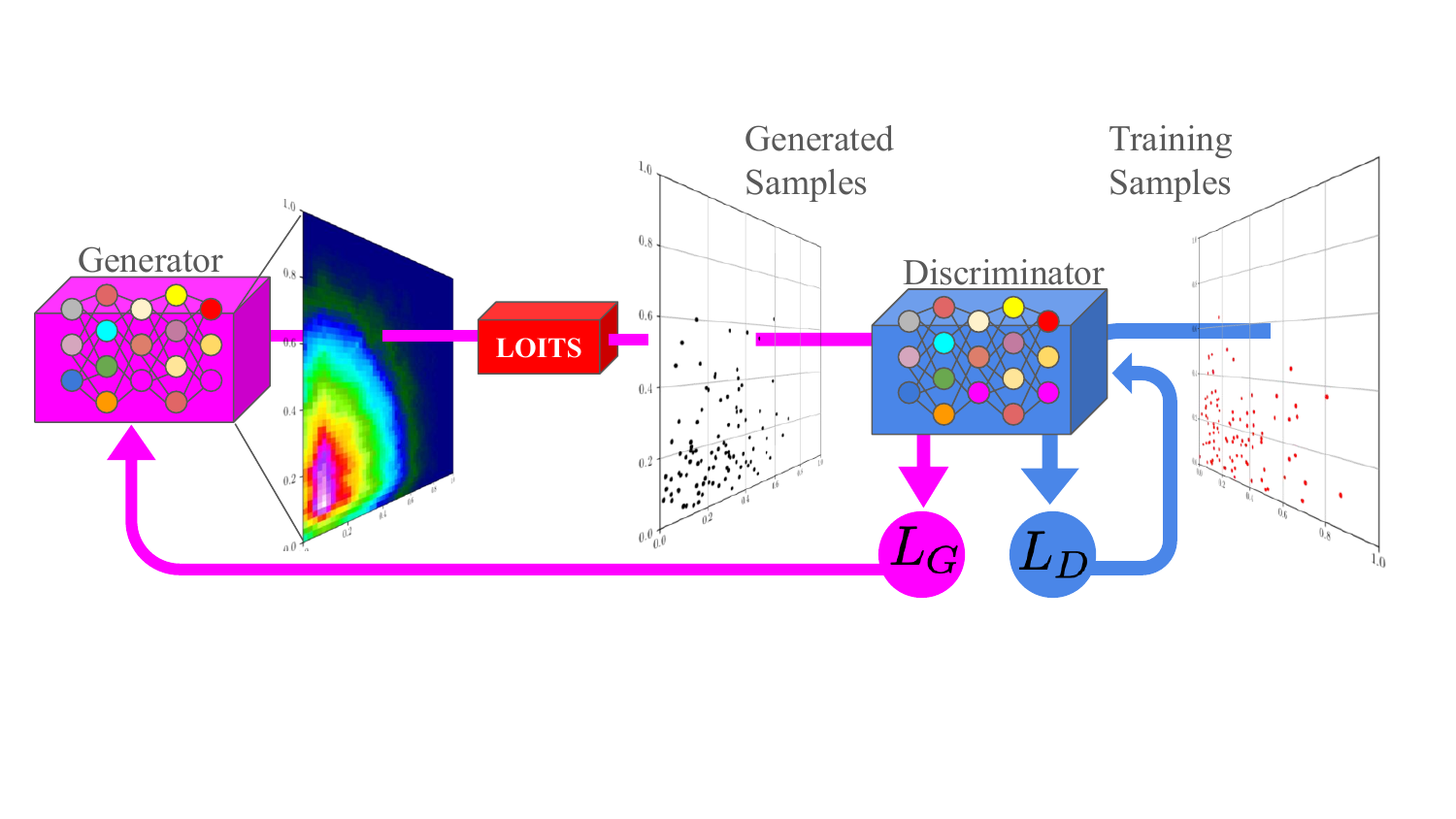}
    \caption{Schematic representation of the GAN setup using the LOITS algorithm.}
    \label{fig:loits_closure}
\end{figure*}

As mentioned earlier, we have assumed equidistant segmentation along each dimension. In principle, there is nothing in the LOITS or LOITS+MH procedure that prevents the use of alternative segmentation strategies. For example, it is possible to define a phase space grid with finer segments in regions where the density exhibits rapid variations and coarser segments in regions where the density is relatively flat. This adaptive segmentation can improve the accuracy of LOITS sampling and increase the acceptance rate in the MH step.

In summary, we have demonstrated that differentiable sampling of a density beyond 1D can be realized using the LOITS algorithm and that its discretization artifacts can be efficiently corrected using a MH accept–reject step. Therefore, the LOITS algorithm provides a differentiable MCMC sampler. Because LOITS-generated proposals already approximate the true target density, the MH acceptance rate remains high, resulting in an efficient sampling procedure with minimal computational overhead. Importantly, when a proposed sample is rejected, the current state is retained in the Markov chain. This step is crucial as it ensures that the sampling trajectory remains continuous and, more importantly, preserves the differentiability of the full sampling process with respect to the underlying model parameters.

Having established a differentiable sampling procedure, the natural question is whether the resulting gradients are informative. That is, given a loss function defined at the sample level, do the gradients with respect to the model parameters provide meaningful directions for optimization? In the next section, we will address this question through a controlled example designed to test the performance of LOITS in a parameter inference task.

\section{Inference using differential sampling and GANs}
\label{s.Test case}

To demonstrate the use of the LOITS algorithm for inference tasks, we consider the following two-dimensional density as the ground truth that we aim to reconstruct
\begin{align}
p(x,y|\phi) = {\cal N} x^{\phi_0} (1-x)^{\phi_1} y^{\phi_2}(1-y)^{\phi_3} (1+\phi_4 x y)\,.
\label{e.model}
\end{align}
Here, ${\cal N}$ is a normalization constant and $\phi$ denotes a fixed set of model parameters used to define the ground truth. The goal of this closure test is to generate training samples $(x,y) \sim p(x,y|\phi)$ and reconstruct the underlying density from a given set of $N$ samples,
\begin{align}
\mathcal{D}_{T}\equiv\{(x_i,y_i)\}_{i=1}^N\sim p(x,y|\phi)\;.    
\end{align}
Here, the subscript ``$T$'' indicates the set of training samples. 
One option for reconstructing the density is to fit the parametric form in Eq.~\eqref{e.model} using unbinned maximum likelihood methods, thereby reducing the task to regressing the parameters $\phi$. Alternatively, the density can be represented as a discretized image, where individual \emph{pixels} of the image must be inferred. In this work, we focus on the latter approach, as it aligns more closely with our intended applications in nuclear imaging. This formulation also allows us to introduce the concept of resolution in the reconstructed image. We emphasize that the LOITS algorithm is applicable to both approaches.

To infer the pixel values of the density image, we use GANs~\cite{NIPS2014_f033ed80}, as illustrated in Fig.~\ref{fig:loits_closure}. The generator network $G$ maps samples from a latent space to a pixelated representation of the density. Mathematically, this can be expressed as 
\begin{align}
p^G_{ij} = {\rm NN}_G(z, \theta_G)\,,
\end{align}
where ${\rm NN}_G$ is a neural network with trainable parameters $\theta_G$, and $z$ denotes latent variables sampled from a fixed prior. The indices $i,j$ label the discretized coordinates $x$ and $y$ of the pixelated density, respectively.

The role of LOITS is to take the discretized density $p^G_{ij}$ as input and generate a set of $N$ samples, denoted by
\begin{align}
    {\cal D}_{G}\equiv\{(x_i^{\theta_G},y_i^{\theta_G})\}_{i=1}^N\sim p^G_{ij}\;.
\end{align}
Here, the subscript ``$G$'' indicates that the samples ${\cal D}_{G}$ are obtained using the density image from the generator network. The full pipeline for the generator can be summarized as
\begin{align}
z \to {\rm NN}_G
\to p^G_{ij}
\to {\rm LOITS}
\to {\cal D}_{G}\,.
\label{eq.Gsim}
\end{align}
To apply LOITS effectively, the pixel resolution of $p^G_{ij}$ needs to exceed the segmentation scale of the LOITS algorithm. For instance, within the blue region in Fig.~\ref{f.2dloits}, a sufficient number of pixels along each orthogonal segment is required to accurately compute the local CDF in each dimension. By construction, the generated samples are differentiable with respect to the pixel values. Moreover, since the pixel values themselves are differentiable with respect to the generator parameters $\theta_G$, one can compute the gradients of the generated samples with respect to $\theta_G$ via the chain rule. For example, the gradient with respect to the $k^{\rm th}$ parameter of the generator can be estimated as 
\begin{align}
\frac{\partial x_{\theta_G}}{\partial \theta^k_G} = \sum_{ij}
\frac{\partial x_{\theta_G}}{\partial p^G_{ij}}
\frac{\partial p^G_{ij}}{\partial \theta^k_G}\;,
\end{align}
and similarly for $y_{\theta_G}$.

To train the generator network $G$, we adopt the standard GAN setup by introducing a discriminator network $D$ that is trained to distinguish generated and real (training) samples. The discriminator is a neural network that we define as
\begin{align}
d = {\rm NN}_D(x, y, \theta_D)\,,
\end{align}
which outputs a scalar classification score. Here, ${\rm NN}_D$ has trainable parameters $\theta_D$, and $(x, y)$ represents an input sample. These samples are drawn either from the training dataset ${\cal D}_{T}$ or from the generated set ${\cal D}_{G}$, depending on the stage of adversarial training. The discriminator's output $d$ serves as a classification score estimating whether a given sample is real or generated.

We follow the GAN training procedure using the binary cross-entropy (BCE) loss to train both networks. In general, the BCE loss is defined as
\begin{align}
L_{\rm BCE}(w, \{d\}) = \frac{1}{N} \sum_i \left[ w \log d_i + (1 - w) \log (1 - d_i) \right]\,,
\end{align}
where $\{d\} = \{d_i = {\rm NN}_D(x_i, y_i, \theta_D) | i = 1, \ldots, N\}$ is the set of discriminator outputs for a batch of samples, and $w \in {0, 1}$ indicates the target label, where 1 corresponds to real samples and 0 to generated samples. The discriminator and generator networks are trained by minimizing the following loss functions:
\begin{align}
L_D &= L_{\rm BCE}(1, \{d\}_T) + L_{\rm BCE}(0, \{d\}_G)\notag\\
L_G &= L_{\rm BCE}(1, \{d\}_G).
\end{align}
Here, $\{d\}_T$ and $\{d\}_G$ denote the discriminator outputs evaluated on the real and generated samples, respectively. The discriminator is trained to correctly classify real vs. generated samples, while the generator is trained to fool the discriminator into classifying generated samples as real.

It is important to emphasize that the generator loss $L_G$ requires the generated samples ${\cal D}_{G}$ to be differentiable with respect to the generator parameters $\theta_G$. This requirement is precisely what the LOITS algorithm is designed to address, enabling end-to-end optimization of the generator via differentiable sample generation.

The advantage of using GANs lies in their ability to infer complex densities or images at the sample level without requiring an explicit definition of the likelihood function. This contrasts with traditional likelihood-based inference methods, such as maximum likelihood estimation or Bayesian inference, which rely on a tractable and often analytically known likelihood function. 

\begin{figure*}[t]
    \centering   
    \includegraphics[width=\textwidth]{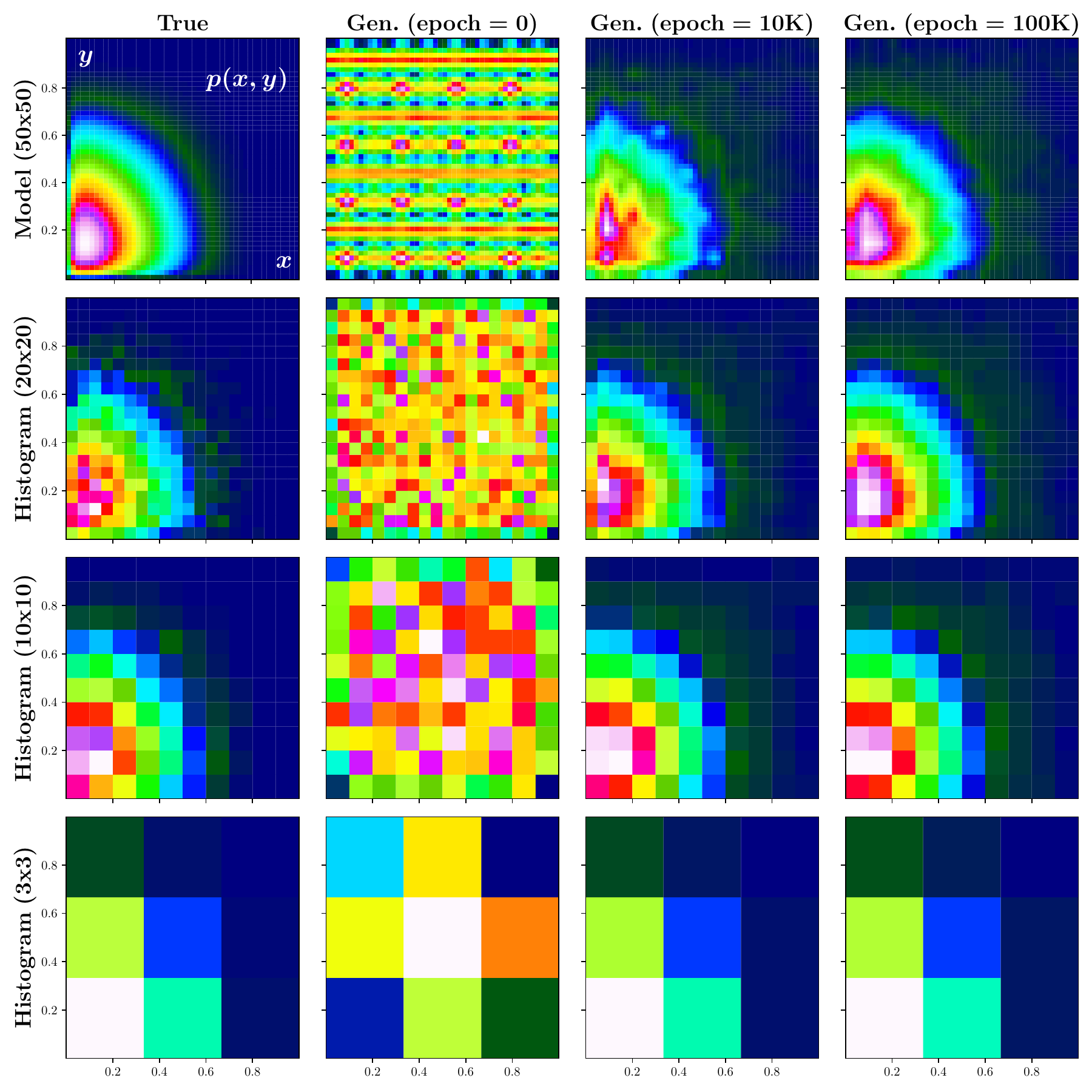}
    \caption{GAN-based image inference of the underlying density supervised on $10\text{k}$ phase space samples. The top row shows the ground truth density image (leftmost panel) alongside GAN-generated images at different stages of training. The subsequent rows display lower-resolution reconstructions obtained by histogramming the phase space samples. As the resolution decreases, the agreement between the GAN-generated density and the ground truth improves, indicating that large-scale features are more robustly captured by the generator.}
    \label{f.2dgan}
\end{figure*}

With these considerations in mind, we now describe the specific setup used for training the GAN:

\begin{itemize}
    \item The generator maps a latent noise vector to an image with $50 \times 50$ pixels, where each pixel corresponds to a discretized coordinate in the $(x, y)$ plane representing the associated density value. The generator is implemented as a fully connected neural network followed by transposed convolutional layers. Specifically, it begins with three linear layers, each containing 100 neurons, followed by four two-dimensional transposed convolutional layers that progressively upsample the features. Each convolutional layer employs 100 filters. The final output is refined using bilinear interpolation to match the target resolution of $50 \times 50$ pixels. All layers use a leaky ReLU activation function with a negative slope of 0.2, except for the final layer, which uses a sigmoid activation to constrain the output values between 0 and 1.

    \item To simplify the GAN training, we average the generator’s output over a batch of 1,000 latent noise samples before passing it to the LOITS algorithm. This averaging produces a smoother density image and improves the efficiency of the training process.
    
    \item The discriminator is trained on batches of $100\text{k}$ samples, drawn either from the real (training) dataset or from the generated set produced via LOITS. It is implemented as a fully connected neural network with four linear layers, each using 128 neurons and leaky ReLU activation. The output layer uses sigmoid activation to produce a binary classification score.
    
    \item Both the generator and discriminator are trained for 100k epochs using 8 NVIDIA T4 GPUs. We employ a distributed training strategy as described in Ref.~\cite{Lersch2024SAGIPS}.
\end{itemize}

\begin{figure*}[t]
    \centering   
    \includegraphics[width=\textwidth]{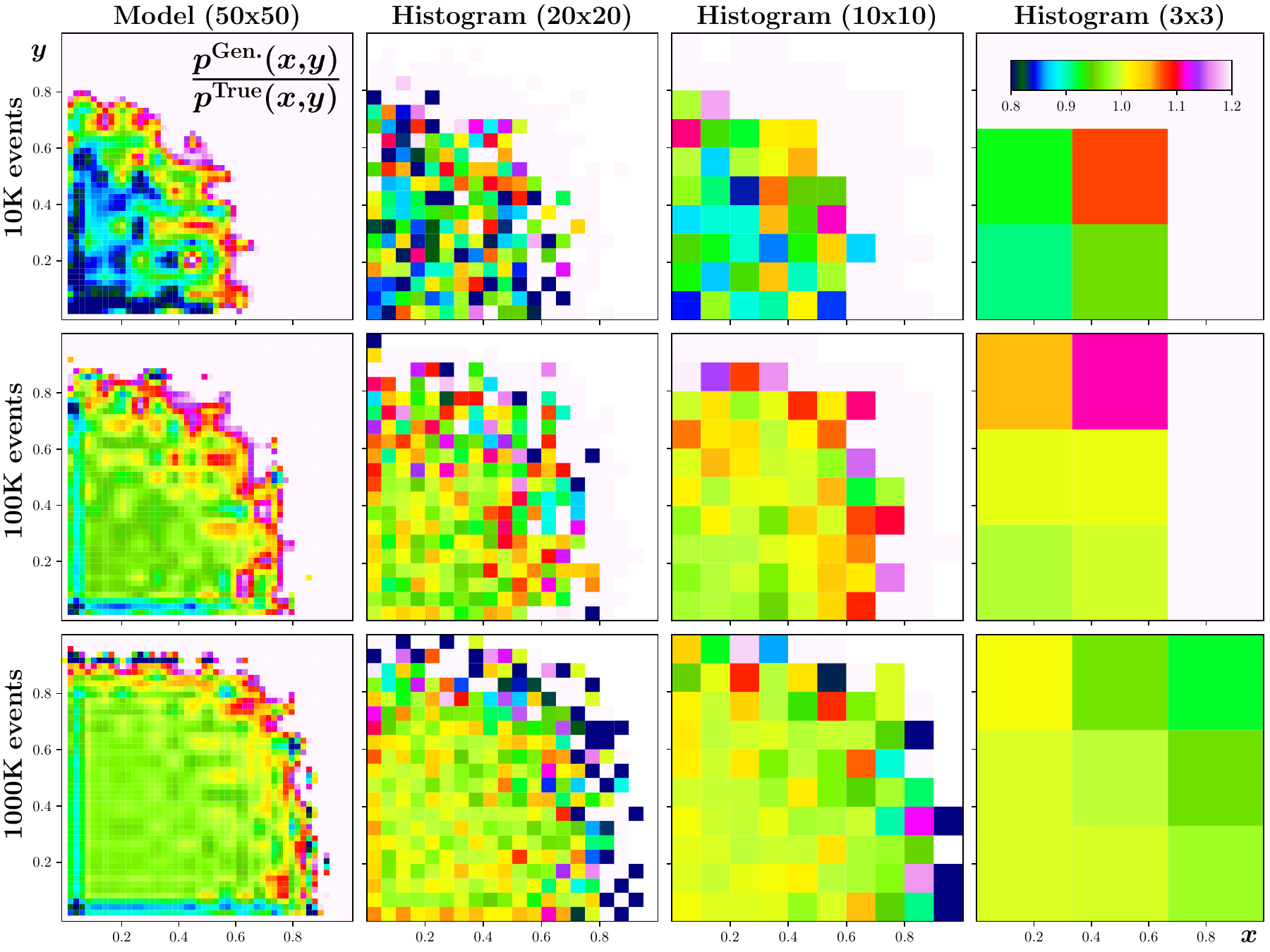}
    \caption{Pixel-wise ratios between GAN-generated and ground truth densities from different training runs, using datasets of varying sizes: $10\text{k}$, $100\text{k}$, and $1000\text{k}$ samples.
    }
    \label{f.2dresol}
\end{figure*}

In Fig.~\ref{f.2dgan}, we present the results of GAN training using $10\text{k}$ phase space samples. The upper leftmost panel shows the ground truth density that the generator is tasked with learning. The remaining three panels in the top row illustrate how the generated image evolves during training. As the number of training epochs increases, the generator progressively improves its output. After approximately $100\text{k}$ epochs, training stabilizes, and the generated image closely approximates the ground truth, although some residual artifacts remain.

To evaluate how well the inferred density matches the ground truth, we estimate the effective resolution of the generator output. While several metrics could be used, we adopt a straightforward approach: histogramming the event-level samples at various binning resolutions. This provides a practical way to assess the scale at which the generated density best reproduces the true underlying distribution. Rows 2–4 of Fig.~\ref{f.2dgan}, analogous to the top row, show these comparisons using progressively coarser binning. As evident from the figure, the agreement improves as the binning becomes coarser, indicating that the generator more reliably captures large-scale structures than fine-grained details. In realistic applications, the effective resolution could be calibrated using simulations as a function of the available event statistics.

The fact that the effective inferred resolution is coarser than the original target image highlights a limitation of the GAN training process. Several factors may contribute to this, including (i) the limited size of the training sample, (ii) the behavior of the LOITS gradients, iii) the architecture and optimization hyperparameters of the neural networks. In Fig.~\ref{f.2dresol}, we isolate the effect of training sample size by repeating the analysis with datasets containing $10\text{k}$, $100\text{k}$, and $1000\text{k}$ samples. The results are shown as pixel-wise ratios between the GAN-generated and ground truth densities, allowing for a direct visual assessment of how reconstruction fidelity improves with increasing data.

The first column of Fig.~\ref{f.2dresol} shows the pixel-wise ratio between the GAN-generated density and the true density. The increasing ratio values close to one across the phase space indicate that, as the number of training samples increases, the GAN more accurately infers the underlying density. Columns 2–4 show lower-resolution comparisons obtained by histogramming the samples, following the same procedure as in Fig.~\ref{f.2dgan}. Across all resolutions, we observe consistent improvements in both the fidelity and effective resolution of the generated density as more data becomes available. These results demonstrate that the LOITS algorithm enables efficient gradient-based optimization for training GANs with up to ${\cal O}(3{\rm M})$ parameters, even in the absence of an explicit likelihood. This allows for direct event-level inference of complex densities using differentiable sample generation.

Although the LOITS gradients vanish exactly at the boundaries of the sampling grid, they remain nonzero away from the edges, which is sufficient to drive the optimization of the GAN. Because samples are drawn from a continuous uniform distribution, see Eq.~(\ref{e.LI}), the probability of generating a point exactly on the grid edge is negligible. While the gradients decay as samples approach the boundaries, their direction, not magnitude, is the crucial quantity for optimization. Gradient-based optimizers such as Adam~\cite{kingma2014adam} can reliably update model parameters as long as the gradients are nonzero within machine precision, regardless of their absolute scale.

\section{Conclusions} 
\label{s.conclusions}

In this work, we introduced an algorithm called Longitudinal Orthogonal Inverse Transform Sampling (LOITS), which constructs a differentiable map between phase space samples and the corresponding multidimensional densities from which those samples are drawn. This algorithm is a key building block enabling, in principle, an end-to-end simulation-based inference pipeline for studying the nonperturbative structure of hadrons, which is encapsulated in various correlation functions such as PDFs, TMDs, and GPDs, directly from event-level data, without relying on surrogate machine learning models.

The algorithm extends inverse transform sampling to multi-dimensional problems and leverages modern auto-differentiation frameworks available in state-of-the-art machine learning libraries to compute gradients efficiently. While the sampling produced by LOITS is approximate, we have demonstrated that its accuracy can be systematically improved by incorporating a Metropolis–Hastings accept-reject step.

To validate the effectiveness of the gradients produced by the LOITS algorithm, we tested it on a simplified two-dimensional problem. Specifically, we employed a GAN as the generative model for density image reconstruction, closely mirroring its intended application to three-dimensional hadron structure studies. The results confirm that LOITS yields accurate and informative gradients for optimizing generative models in a physics-informed, likelihood-free inference setting. While our tests focused on GANs, the LOITS algorithm is broadly applicable and can be integrated with other machine learning approaches, such as normalizing flows, or even applied to classical unbinned likelihood analyses.

In addition, the imaging-based approach has enabled us to quantify the resolution that GAN models can achieve when trained on event-level information. Our results confirm that the fidelity of the reconstructed density image improves with increasing amounts of training data, indicating a direct connection between sample size and learnable resolution. This provides a valuable tool for uncertainty quantification in the emerging era of femtoscale imaging, where precise control over resolution is essential for extracting QCFs from observational data. 

An end-to-end simulation-based inference framework also requires the availability of differentiable detector simulations. Currently, simulators such as \textsc{Geant}4~\cite{Agostinelli:2002hh} are not differentiable; however, surrogates based on generative modeling are becoming available~\cite{Krause:2024avx} and can be integrated into our proposed framework. Alternatively, sample-level unfolding techniques provide another path forward, and the LOITS algorithm can be incorporated into inference pipelines that utilize unfolded data. 

While the development of the LOITS algorithm was motivated by challenges in hadron structure studies, particularly inverse problems involving multidimensional distributions such as PDFs, TMDs, and GPDs, we anticipate that its applicability extends well beyond this domain. In particular, LOITS may offer significant value in a broad class of optimization problems that require a differentiable mapping between stochastic observables and the underlying fundamental degrees of freedom.

\begin{acknowledgments}
This work was supported by the DOE contract No. DE-AC05-06OR23177, under which Jefferson Science Associates, LLC operates Jefferson Lab; the U.S. Department of Energy, Office of Science, Office of Nuclear Physics, contract no. DE-AC02-06CH11357; the Scientific Discovery through Advanced Computing (SciDAC) award Femtoscale Imaging of Nuclei using Exascale Platforms. FR was supported by the DOE Quark-Gluon Tomography (QGT) Topical Collaboration under contract No. DE-SC0023646 and by the DOE, Office of Science, Office of Nuclear Physics, Early Career Program under contract No. DE-SC0024358 and DE-SC0025881. NS was supported by the DOE, Office of Science, Office of Nuclear Physics in the Early Career Program.
\end{acknowledgments}

\appendix

\section{Double half-moon distribution~\label{app:halfmoon}}

For completeness, we provide here the analytic expressions for the double half-moon (DHM) test distribution used in Section~\ref{s.loits}. The DHM distribution is constructed from two Gaussian rings of the form:
\begin{align}
f_{\rm ring}(x,y;\boldsymbol{r}^i)
=\exp\left(-\frac{1}{2}\left(\frac{r(x,y;\boldsymbol{r}^i)-R}{\sigma}\right)^2\right)\; ,    
\end{align}
where $R$ is the mean ring radius, and $r(x,y;\boldsymbol{r}^i)$ denotes the radial distance from the point $(x,y)$ to the center $\boldsymbol{r}^i = (r^i_x, r^i_y)$. We take the centers of the two Gaussian rings as
$    \boldsymbol{r}^{\pm} 
    =\left(\frac{1}{2} \mp \Delta x,\frac{1}{2} \pm \Delta y\right)\;,
$
and restrict each ring to one side of the plane using soft gating functions:
\begin{align}
    {\rm gate}^+(y) &= \frac{1}{2}
    \left(1 + \tanh\left(\frac{y - r^+_y}{w}\right)\right) \;, \\
    {\rm gate}^-(y) &= \frac{1}{2}
    \left(1 + \tanh\left(\frac{r^-_y - y}{w}\right)\right) \;.
\end{align}
The unnormalized DHM density is then defined as
\begin{align}
   p_{\rm DHM}(x,y) 
   = \sum_{i=+,-} f_{\rm ring}(x,y;\boldsymbol{r}^{i})\,{\rm gate}^i(y) \;.
\end{align}
and we have set the DHM  parameters of as
$\Delta x = 0.1$, $\Delta y = 0.1$, $w = 0.1$, $R = 0.2$, $\sigma = 0.02$.

\bibliographystyle{utphys}
\bibliography{references}


\end{document}